# Pinching and Probing of Polygonal Grain Boundaries

N. Sarkar[1*], P.R. Bandaru,[1,3*] R.C. Dynes[2,3*]

[1]Department of Mechanical Engineering, [2]Department of Physics,

[3]Program in Materials Science,

University of California San Diego, La Jolla, California 92093-0411, USA

Abstract: In this study, sub-angstrom spatial resolution is achieved in mapping and spectroscopy of atoms and bonds within polygonal grain boundaries (GBs) of graphite using Scanning Tunneling Microscopy (STM). Robust van Hove singularities (VHS) are observed in addition to edge states under ambient conditions. The bias-dependent nature of these states reveals metallic traits of GB, through the charge accumulation and dissipation of localized electronic states. Utilizing a surface elastic deformation technique induced by STM tip allows "pico-pinching" of the GB, providing insights into its mechanical strength as well as in-situ strain-induced modification of their unique spectroscopy, revealing a tendency toward flattening of the electronic energy band dispersion. An initial atomic-level experimental technique of probing spin-polarized magnetic states is demonstrated, suggesting different densities for spin-up and spin-down states within a spin-degenerate band structure potentially applicable in spin transport or quantum spin sensing.

Main: Confined within the atomic sheets of two-dimensional (2D) materials, are one-dimensional (1D) grain boundary (GB) structures with distinctive and periodic stitching patterns (1,2) that can induce the emergence of intricate correlated phases(3) and electronic states, such as van Hove singularities (VHS) (4,5), edge states (6–8), strain-induced electronic flat bands (9,10), p-n-p junctions (11), metallicity(2,12), ferromagnetism(13,14), and more, as elucidated in the present work utilizing Scanning Tunneling Microscopy (STM) on highly oriented pyrolytic graphite (HoPG). Such GBs with predictable properties could be potentially applicable in transistors (11,15), electronic waveguides (16), veselago lenses (17), single molecule detection sensors (18,19), solar cells (20), klein tunneling effect (21,22), etc. The GBs of hexagonal HoPG sheet are comprised of pentagons (5), heptagons (7) and octagons (8) arranged in periodic patterns depending on grain orientation angle ($\theta_{orient}$) and substrate-graphite/graphene interactions during growth (23,24). For instance, strongly interacting SiC (25) and Ni substrates (26) through chemical vapor deposition (CVD) growth allows the formation of Stone Wall point defects and 5-8-5 GBs (with $\theta_{orient}$ = 10°) respectively but weakly interacting substrates like Cu in CVD (27) or HoPG production (28) statistically favors zigzag patterned 5-7 GBs (with $\theta_{orient}$ = 29°) as shown in Fig. **1**. While atomic configurations of GB can be revealed with high resolution electron microscopy, a comprehensive understanding, incorporating the interplay of the mechanical, electronic, or magnetic properties on atomic scale can only be probed by scanning probe microscopy (SPM) (13,29–32). Often, the bright regions in STM topography maps may be obscured by potential minor electronic contributions originating from the high density of localized electronic states at GB sites, where the obscured atomic configuration then has been subsequently inferred through simulated local density of states (LDOS) map calculations (4,5,33).

The 3D and 2D STM topography of such a periodic GB is shown in Fig. **1(a-b),** illustrated in Fig. **1(c)** which shows a periodicity of 1.36 nm as exhibited in Fig. **1(d)** that includes two 5-7 pairs denoted by (1,0) and (0,1) and $\theta_{orient}$ ~ 29° for the GB is estimated through a fast Fourier transform map in Fig. **1(e)**. In this paper, we present zigzag-patterned images of the 5-7 GB with angstrom-level spatial resolution in spectroscopic probing of LDOS at atom/bond locations on pentagons or heptagons and nearby hexagons. The origin of these bias-specific conductance states is revealed as VHS at high biases, also previously recognized (4), along with the emergence of edge states at low biases. Experimentally observed edge states in previous graphene nanoribbon (GNR) studies (7) demonstrate that zigzag edges, in contrast to armchair edges, exhibit a high DOS at zero bias, resulting in a flat band due to state localization at zigzag kinks, as illustrated in Fig. **1(f)**. Similar edge effects on zigzag kinks have been observed on graphite step edges with alternating partial zigzag and armchair edges, also resembling the geometry of a 5-7 GB edge shown in Fig. **1(d)** which illustrates when illustratively de-stitched of the GB and revealing anti-parallel equal segments (~1 nm) of alternating zigzag and armchair edges (8).

The mechanical strength of polycrystalline graphite sheet is weakened by its GBs, from atomic force microscopy (AFM) based mechanical tests (31,32). GBs also tend to be preferential sites for step edge creation during exfoliation or cleavage. A 5-7 pair i.e. (1,0) or (0,1) is energetically more stable than a single 5 or 7 (33). The configuration, where (1,0) and (0,1) dislocations are symmetrically oriented in a linear zigzag chain, could further introduce mechanical stability (30). Often the GB is paired (34) with a nano-wrinkle in parallel or another GB in anti-parallel for strain relaxation. Further, the in-plane elastic energy of these dislocations causes out-of-plane atomic 'buckling' of the GB which further relieves their formation energies and screens the in-plane strain field (33).

In Fig. **2(a)**, a linear chain of 5-7 pairs comprising the GB exhibited a measured buckling amplitude of ~0.35 nm, aligning closely with the theoretical prediction (33) of 0.3 nm for a single 5-7 pair in a free-standing graphene. The atomic sheets of HoPG are also nearly free-standing as they are weakly coupled by vdW forces, enabling the 'soft' HoPG surface (35,36) to be easily deformed out-of-plane under the influence of STM tip-induced atomic forces at lower tip-sample gap distances (or lower gap resistance $R_{gap}= V_{sample}/I_{tip}$) as illustrated in Fig. **2(b)**. The elastic deformation extent is shown as a function of $R_{gap}$ for graphite grains and GB in Fig. **2(c)**. Prior tip-induced deformation study (9,36) on graphite moiré pattern has revealed larger(\smaller) deformation extent for strongly(\weakly) bonded vdW stacking regions of the moiré pattern which can be used as methodology for comparing relative interlayer vdW strength of other features like GB. The buckled GB is estimated to be as strongly bonded to the underlying layer as the Bernal stacked graphite (36).

The atomically flat and clean surface of cleaved HoPG is also used as height calibration standard in STM studies which can also be an ideal substrate for reliable structural and spectroscopic investigations on graphite GB, including strain induced effects (36–39). The topography across various sites of the buckled GB (see Supplemental Material Sec. S1) reveals that only a single GB atom is deformed out-of-plane, also illustrated in Fig. **1(c)** (shown by light green atoms). Previous simulation mappings have depicted higher electronic LDOS around these deformed GB atoms, as depicted in Fig. **3(a)** where there is electronic cloud (greenish yellow) localized around the pentagons, indicating higher charge density near the pentagons (or lower near the heptagons), resulting in a negative (or positive) charge distribution, also denoted as self-doping (4,5,33).

This charge distribution around the GB region can be probed by conductance spectroscopy of electronic states gradually across the polygons, along a path (yellow arrow), Pentagon → Heptagon → Hexagon, as shown in Fig. **3(a)**. Of significance in this work is that the local conductance spectroscopy is captured with exceptional spatial determinations spanning just one-tenth of a nm. Here, unique bias-specific peaks are observed for the polygons (Fig. **3(b)**). For instance, on heptagons, distinct VHS peaks are notably exhibited at +/- ~400 mV from the high conductance and expected large DOS with flat-band characteristics. This is in agreement with the calculated band structure (Fig. **5(c)**) where the VHS states at -400(/+400) mV correspond to filled (/empty) bands (4,33). In addition, as the tip traverses from pentagons to heptagons, a novel set of spectroscopic features has emerged, revealing a gradual transition from a single +400 mV peak to three distinct peaks. These conductance peaks closely align with the calculated LDOS in Fig. **5(a)**. The VHS states on 5's and 7's gradually disappear as the neighboring hexagons are encountered and a pair of peaks at low bias are observed, denoted as low bias states (LBS: LBS#1 and LBS#2) in Fig. **3(b)** (bottom). The conductance curve on hexagons also exhibits a minima, known as the Dirac point, positioned at +183 mV as denoted by a black triangle.

From 6 → 5 → 7, a shift in the conductance minima, relative to zero bias, is observed within the yellow dotted box in Fig. **3(b)**, signifying local electronic doping from the bulk. The minima at 183 mV (off the GB) indicates p-doped grains which gradually transition to –60 mV (on the GB) indicating n-doped GB, in agreement with previous work ((40). Besides the doping-induced minima shift, we identified an alternative mechanism for further shifting the conductance curve minima of heptagons and pentagons when subjected to tip-induced local strain, as depicted in Fig. **4(a)** and Fig. **4(b)** respectively. *In-situ* spectroscopy as a function of deformation demonstrates not only widening of the conductance peaks but also a shift in the conductance minima for heptagons and pentagons with decreasing $R_{gap}$ [Fig. **4**, bottom to top]. The deformation of the GB surface at lower $R_{gap}$ indicated a tendency towards increased electrical conductance, which may be associated with a higher density of electronic states, and the consequent flattening of the electronic energy band dispersion. Additionally, the deformation-induced minima shift indicates a possibility towards formation of asymmetric tunneling barriers at low gap distances, potentially accounting for a minor shift in the conductance minima (up to ~100mV), in agreement with the Brinkman, Dynes and Rowell (BDR) model (41).

The spatial distribution of the LBS and VHS states can be observed with bias-dependent imaging as shown in Fig. **5(d-e-f-g)**. Fig. **5(g)** indicates that the VHS states at 400 mV associated with GB 'empty bands' (in Fig. **5(c)**) are slightly delocalized and spread across the GB by ~1 nm. However, the distribution of LBS appears to be localized only along the GB with the increase in $V_{sample}$ from 25 mV → 150 mV → 240 mV in Fig. **5(d)** → **(e)** → **(f)**. Bias-dependent imaging observations are similar at negative $V_{sample}$ too (Refer Supplementary S2). Consistent and reversible bias-dependent imaging on atomic sites, coupled with the absence of charge retainment at those sites within +/- 500 mV, suggests the metallic nature of the GB. This high transparency to charge carriers was theoretically predicted (2,42) only for symmetric GBs spanning grains with identical translational vectors, which is also the case for this zigzag GB.

Several interpretations for LBS peaks have been proposed (5) such as possible moiré-based interlayer interactions with underlying twisted graphite sheet, but experimental evidence of LBS exists for single-layer graphene GBs with no twist (27). An alternative explanation posits phonon-mediated inelastic tunneling of electrons, causing similar gapped features, observed within 2.4 nm of GB (7,43). This contrasts with our observed spatially confined LBS to GB sites only in Fig. **5(d-f)** where Fig. **5(d)** shows that LBS #1 state is confined only around (0,1) dislocation pair, as also depicted in Fig. **5(b)**. With the increase in $V_{sample}$, LBS gets delocalized along the 1D GB (refer Fig. **5(f)**), as also illustrated in simulated LDOS map in Fig. **3(a)**, where the electronic cloud is centered around the light green atoms. These atoms are also revealed as edge carbon atoms located at the kinks of partial zigzag edges in Fig. **1(d)** that electronically exhibit low-bias edge states, also indicated in the calculated band structure of Fig. **5(c)** (green curve). The flat band of edge states suggests that a single conductance peak at low bias should be expected at GB vicinity, but we observe a pair of gapped LBS peaks (with gap width (Δ) ~ 250 mV). The partial zigzag edges of the GB are separated by an armchair to limit the electronic interaction between the edge states of two partial zigzag edges. However, finite electron-electron interaction can spin-polarize these edge states that split the low-energy flat band single peak into a pair of peaks (with Δ ~ 65 mV), causing ferromagnetic (/antiferromagnetic) correlations along (/across) the GB (7). However, in this model, the out-of-plane buckling of the zigzag edge atoms, possibly inducing pseudo-field effects, was excluded which could account for the Δ differences (44).

Indeed, room-temperature ferromagnetism has been observed on disordered HoPG GB (13). We have also demonstrated a technique of probing spin-polarized magnetic states on a typical 1D GB by application of switchable and reversible external magnetic fields on the sample through a current carrying coil wrapped around the tip as illustrated in Fig. **6(a)**. The magnetic field can be switched OFF in **(b)** and ON in **(c)**, further magnified in **(d)**, and field-reversed in **(e)**, revealing the spatial distribution of spin-up and spin-down states. This study provides the initial atomic-level experimental evidence that localized electronic states at atomic sites respond to magnetic field reversals, suggesting different densities for spin-up and spin-down states within a spin-degenerate band structure potentially applicable in spin transport or quantum spin sensing. This aligns with the unresolved origin of LBS to possibly spin-polarized edge states as also expected in vicinity of GNR edges.

## **Methods:**

## Materials

HoPG of low ZYH grade from Advanced Ceramics, Inc with micron sized grains is synthesized at ~3000°C which favors the relaxation of the grain boundaries into periodic defects in contrast to aperiodic GBs in CVD graphene grown at 1000°C.

## Electrical Measurements

A custom-built scanning tunneling microscope (STM) with a RHK controller at room temperature and atmospheric pressure was used. The imaging was done at a nominal sample bias of 100 mV and tunneling current of 10nA. The topography measurements were performed in standard constant current mode at ~0.5 Hz scanning frequency using a mechanically snipped Pt/Ir tip. For surface deformation experiments, the tip was brought closer to the sample by varying the tunneling currents from 10 nA to 100 nA. Tunneling spectroscopy measurements utilized a lock-in modulation of 3mV at 5kHz to the sample bias which was swept from –400 mV to 500 mV.

## Calibration and Data analysis

The reported topographic length scales in-plane (x, y) and out-of-plane (z) were calibrated through atomic lattice constants of graphite through atomic imaging of the surface and its monatomic step. The atomic corrugation estimation were averaged over ten atomic unit cells. All the imaging and deformation measurements were done using the same tip. All images were measured with $V_{sample}$ is 100 mV.

## Data availability statement

The experimental data and its analysis in the paper and/or in the supplementary information is sufficient to support our conclusions. Additional data can be made available on request.

## Acknowledgements

This work was supported by AFOSR Grant (FA9550-15-1-0218) and Army Research Office (AROW911NF-21-1-0041). The authors thank Michael Rezin and Rich Barber for the technical

**References:**

1.  Huang PY, Ruiz-Vargas CS, Van Der Zande AM, Whitney WS, Levendorf MP, Kevek JW, et al. Grains and grain boundaries in single-layer graphene atomic patchwork quilts. Nature. 2011;469(7330).

2.  Yazyev O V., Louie SG. Electronic transport in polycrystalline graphene. Nat Mater. 2010;9(10).

3.  Hsieh K, Kochat V, Biswas T, Tiwary CS, Mishra A, Ramalingam G, et al. Spontaneous Time-Reversal Symmetry Breaking at Individual Grain Boundaries in Graphene. Phys Rev Lett. 2021;126(20).

4.  Ma C, Sun H, Zhao Y, Li B, Li Q, Zhao A, et al. Evidence of van Hove singularities in ordered grain boundaries of graphene. Phys Rev Lett. 2014 Jun 6;112(22).

5.  Luican-Mayer A, Barrios-Vargas JE, Falkenberg JT, Autès G, Cummings AW, Soriano D, et al. Localized electronic states at grain boundaries on the surface of graphene and graphite.

6.  Fujita M, Katsunori W, Nakada K, Kusakabe K. Peculiar Localized State at Zigzag Graphite Edge. J Physical Soc Japan [Internet]. 1996 [cited 2023 Sep 16];65(7):1920–3. Available from: https://www.jstage.jst.go.jp/article/jpsj/65/7/65_7_1920/_article/-char/ja/

7.  Tao C, Jiao L, Yazyev O V., Chen YC, Feng J, Zhang X, et al. Spatially resolving edge states of chiral graphene nanoribbons. Nat Phys. 2011;7(8):616–20.

8.  Kobayashi Y, Fukui KI, Enoki T, Kusakabe K. Edge state on hydrogen-terminated graphite edges investigated by scanning tunneling microscopy. Phys Rev B Condens Matter Mater Phys. 2006;73(12).

9.  Sarkar N, Bandaru PR, Dynes RC. Characteristic nanoscale deformation on a large-area coherent graphite moiré pattern. Phys Rev B. 2023;107(16).

10. Mesple F, Missaoui A, Cea T, Huder L, Guinea F, Trambly De Laissardière G, et al. Heterostrain Determines Flat Bands in Magic-Angle Twisted Graphene Layers. Phys Rev Lett. 2021;127(12).

11. Lahiri J, Lin Y, Bozkurt P, Oleynik II, Batzill M. An extended defect in graphene as a metallic wire. Nat Nanotechnol. 2010;5(5):326–9.

12. Ihnatsenka S, Zozoulenko I V. Electron interaction, charging, and screening at grain boundaries in graphene. Phys Rev B Condens Matter Mater Phys. 2013;88(8).

13. Červenka J, Katsnelson MI, Flipse CFJ. Room-temperature ferromagnetism in graphite driven by two-dimensional networks of pointdefects. Nat Phys. 2009;5(11).

14. Zhang Z, Zou X, Crespi VH, Yakobson BI. Intrinsic magnetism of grain boundaries in two-dimensional metal dichalcogenides. ACS Nano. 2013;7(12).

15. Jiménez D, Cummings AW, Chaves F, Van Tuan D, Kotakoski J, Roche S. Impact of graphene polycrystallinity on the performance of graphene field-effect transistors. Appl Phys Lett. 2014;104(4).

16. Márk GI, Vancsó P, Lambin P, Hwang C, Biró LP. Forming electronic waveguides from graphene grain boundaries. J Nanophotonics. 2012;6(1).

17. Cheianov V V., Fal'ko V, Altshuler BL. The focusing of electron flow and a veselago lens in graphene p-n junctions. Science (1979). 2007;315(5816).

18. Ban DK, Liu Y, Wang Z, Ramachandran S, Sarkar N, Shi Z, et al. Direct DNA Methylation Profiling with an Electric Biosensor. ACS Nano. 2020;14(6).

19. Yasaei P, Kumar B, Hantehzadeh R, Kayyalha M, Baskin A, Repnin N, et al. Chemical sensing with switchable transport channels in graphene grain boundaries. Nat Commun. 2014;5.

20. Fang X, Ding J, Yuan N, Sun P, Lv M, Ding G, et al. Graphene quantum dot incorporated perovskite films: Passivating grain boundaries and facilitating electron extraction. Physical Chemistry Chemical Physics. 2017;19(8).

21. Stander N, Huard B, Goldhaber-Gordon D. Evidence for Klein tunneling in graphene p-n junctions. Phys Rev Lett. 2009 Jan 12;102(2).

22. Katsnelson MI, Novoselov KS, Geim AK. Chiral tunnelling and the Klein paradox in graphene. Nat Phys. 2006;2(9).

23. Biró LP, Lambin P. Grain boundaries in graphene grown by chemical vapor deposition. New J Phys. 2013 Mar;15.

24. Yu Q, Jauregui LA, Wu W, Colby R, Tian J, Su Z, et al. Control and characterization of individual grains and grain boundaries in graphene grown by chemical vapour deposition. Nat Mater. 2011;10(6).

25. Tison Y, Lagoute J, Repain V, Chacon C, Girard Y, Joucken F, et al. Grain boundaries in graphene on SiC(0001⁻) substrate. Nano Lett. 2014;14(11).

26. Zhang Y, Zhang L, Zhou C. Review of chemical vapor deposition of graphene and related applications. Acc Chem Res. 2013;46(10).

27. Yang B, Xu H, Lu J, Loh KP. Periodic grain boundaries formed by thermal reconstruction of polycrystalline graphene film. J Am Chem Soc. 2014 Aug 27;136(34):12041–6.

28. Simonis P, Goffaux C, Thiry PA, Biro LP, Lambin P, Meunier V. STM study of a grain boundary in graphite. Surf Sci. 2002;511(1–3).

29. Koepke JC, Wood JD, Estrada D, Ong ZY, He KT, Pop E, et al. Atomic-scale evidence for potential barriers and strong carrier scattering at graphene grain boundaries: A scanning tunneling microscopy study. ACS Nano. 2013;7(1).

30. Liu Y, Yakobson BI. Cones, pringles, and grain boundary landscapes in graphene topology. Nano Lett. 2010;10(6).

31. Rasool HI, Ophus C, Klug WS, Zettl A, Gimzewski JK. Measurement of the intrinsic strength of crystalline and polycrystalline graphene. Nat Commun. 2013;4.

32. Grantab R, Shenoy VB, Ruoff RS. Anomalous strength characteristics of tilt grain boundaries in graphene. Science (1979). 2010;330(6006).

33. Yazyev O V., Louie SG. Topological defects in graphene: Dislocations and grain boundaries. Phys Rev B Condens Matter Mater Phys. 2010 May 14;81(19).

34. Capasso A, Placidi E, Zhan HF, Perfetto E, Bell JM, Gu YT, et al. Graphene ripples generated by grain boundaries in highly ordered pyrolytic graphite. Carbon N Y. 2014;68.

35. Soler JM, Baro AM, Garcia N, Rohrer H. Interatomic forces in scanning tunneling microscopy: giant corrugations of the graphite surface. Phys Rev Lett. 1986;57(4):444.

36. SARKAR N, Bandaru PR, Dynes R. Probing interlayer van der Waals strengths of two-dimensional surfaces and defects, through STM tip-induced elastic deformations. Nanotechnology [Internet]. 2023; Available from: http://iopscience.iop.org/article/10.1088/1361-6528/acb442

37. Szendro M, Pálinkás A, Süle P, Osváth Z. Anisotropic strain effects in small-twist-angle graphene on graphite. Phys Rev B. 2019;100(12).

38. Batra IP, Garcia N, Rohrer H, Salemink H, Stoll E, Ciraci S. A study of graphite surface with STM and electronic structure calculations. Surf Sci. 1987;181(1–2):126–38.

39. Meyer JC, Geim AK, Katsnelson MI, Novoselov KS, Obergfell D, Roth S, et al. On the roughness of single-and bi-layer graphene membranes. Solid State Commun. 2007;143(1–2):101–9.

40. Tapasztó L, Nemes-Incze P, Dobrik G, Jae Yoo K, Hwang C, Biró LP. Mapping the electronic properties of individual graphene grain boundaries. Appl Phys Lett. 2012 Jan 30;100(5).

41. Brinkman WF, Dynes RC, Rowell JM. Tunneling conductance of asymmetrical barriers. J Appl Phys. 1970;41(5).

42. Majee AK, Foss CJ, Aksamija Z. Impact of Mismatch Angle on Electronic Transport Across Grain Boundaries and Interfaces in 2D Materials. Sci Rep. 2017;7(1).

43. Zhang Y, Brar VW, Wang F, Girit C, Yayon Y, Panlasigui M, et al. Giant phonon-induced conductance in scanning tunnelling spectroscopy of gate-tunable graphene. Nat Phys. 2008;4(8):627–30.

44. Levy N, Burke SA, Meaker KL, Panlasigui M, Zettl A, Guinea F, et al. Strain-induced pseudo-magnetic fields greater than 300 tesla in graphene nanobubbles. Science (1979). 2010;329(5991).

**Figures:**

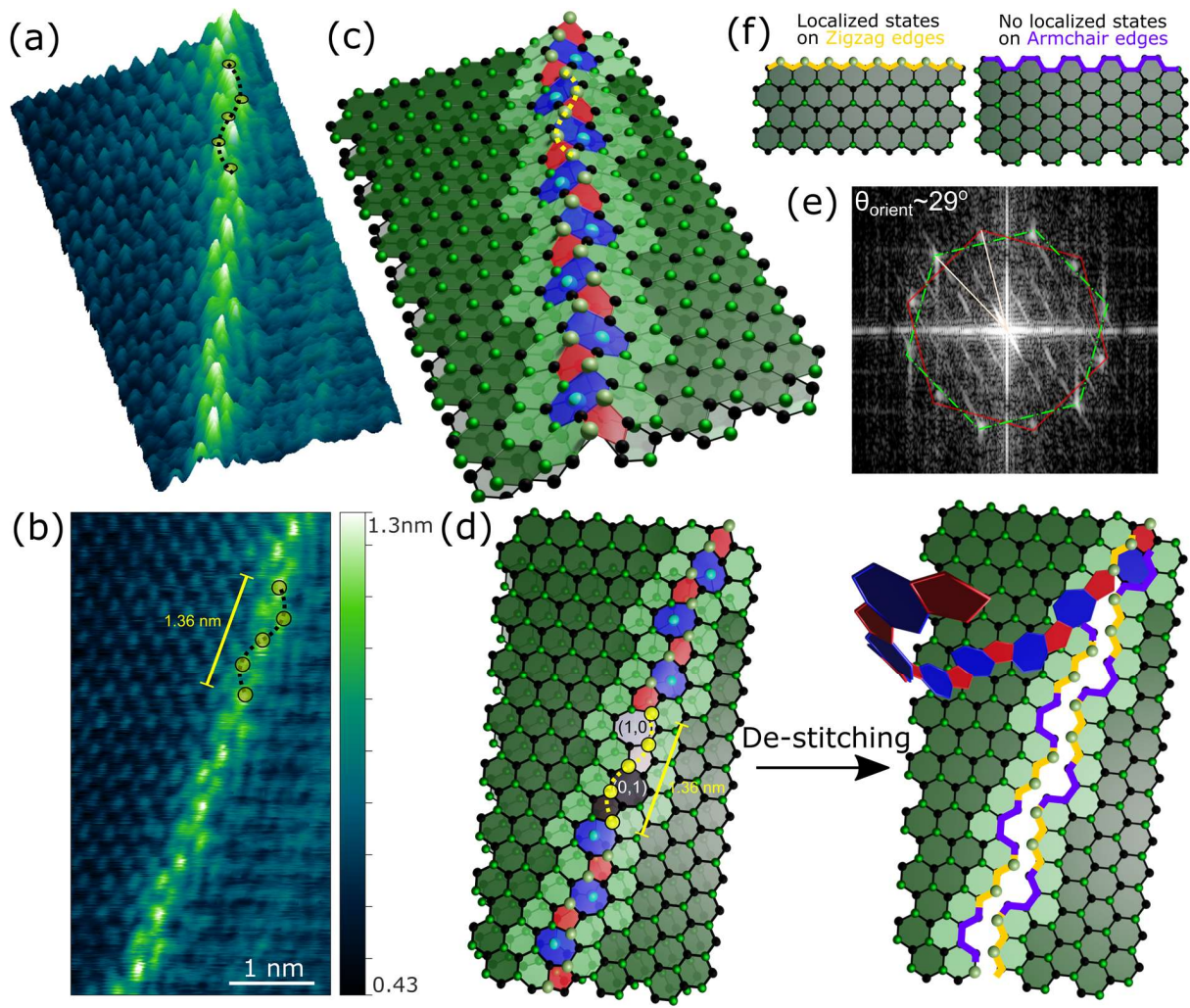

Figure 1: **Zigzag grain boundary on HoPG. (a,b)** 3D and 2D atomically resolved topography map of zigzag grain boundary comprising of pentagons and septagons as illustrated in **(c).** which shows a periodicity of 1.36 nm as exhibited in **(d)** that includes two 5-7 pairs denoted by (1,0) and (0,1) and $\theta_{orient} \sim 29°$ for the GB is estimated through a fast Fourier transform map in **(e)**. De-stitching the GB in **(d)** reveals alternating zigzag and armchair edges along the GB where localized electronic edge states exist at zigzag kinks (shown by light green atoms), also shown in **(f)**.

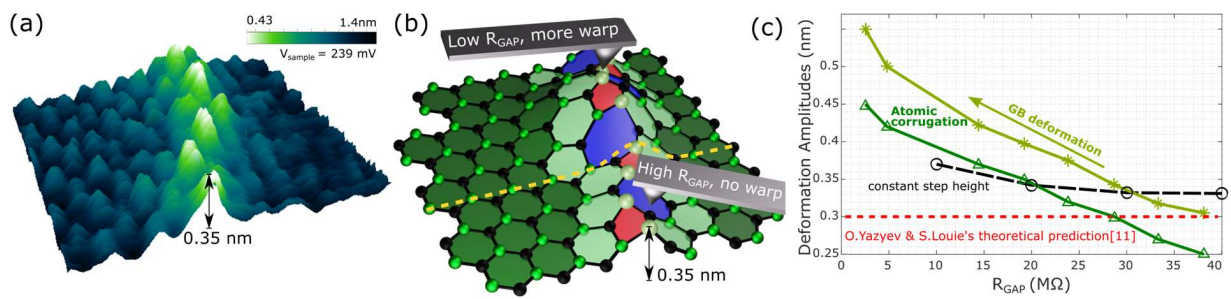

Figure 2: **STM tip induced deformation of GB. (a)** 3D atomically resolved topography showing the out-of-plane buckling of the GB as theoretically predicted (33). **(b)** Schematic showing surface deformation of GB by STM tip through large(/small) atomic forces monitored through large(/small) tunneling currents at smaller(/larger) tip-sample distance. **(c)** The deformation extent of the GB atomic amplitudes and the grains plotted as a function of $R_{gap}$.

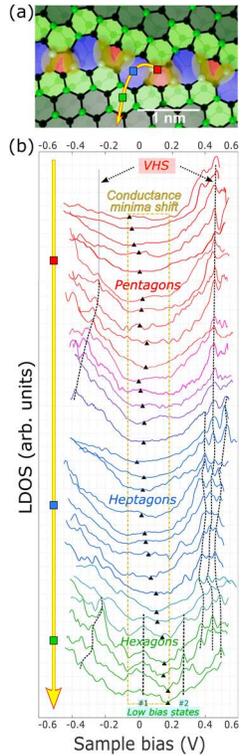

Figure 3: **Electrical conductance mapping across and along the polygons. (a)** Illustrated LDOS map on pentagons and heptagons (adapted from (4,33) showing electronic cloud (greenish yellow) localized around the pentagons, indicating higher charge density near the pentagons (or lower near the heptagons).(4,5,33) **(b)** Moving the STM tip along the *yellow* arrow in **(a)** shows distinctive spectroscopy high bias peaks at +/- ~400 mV on pentagons evolving into triplet peaks on heptagons in contrast to low bias peaks on hexagons at GB vicinity.

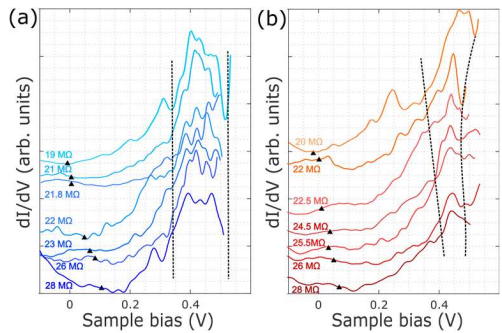

Figure 4: **Conductance peak modifications with surface deformation.** In-situ spectroscopy as a function of deformation demonstrates widening of the conductance peaks and a shift in the conductance minima for heptagons and pentagons with decreasing $R_{gap}$ from bottom to top.

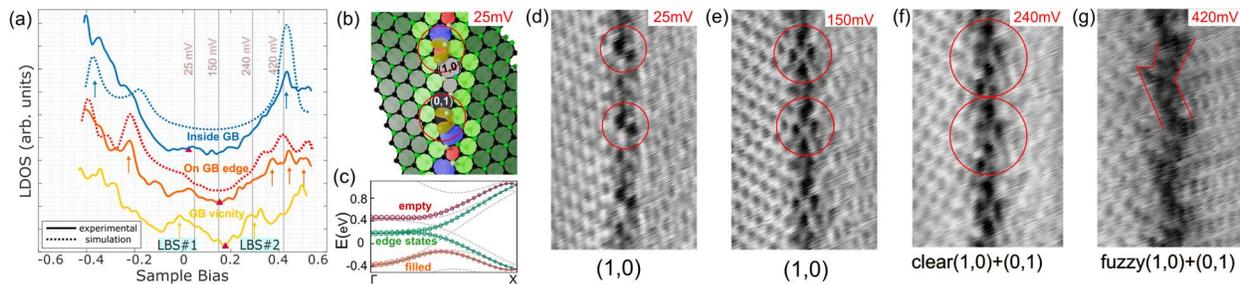

Figure 5: **Bias dependent contant current topography mapping. (a)** The simulated LDOS (adapted from (33)) closely aligns with the experimental conductance. **(b)** Illustrated LDOS map at $V_{sample}$ ~ 25 mV corresponding to experimental STM map in **(d)**. **(c)** Calculated band structure where the VHS states at -400(/+400) mV corresponds to filled (/empty) bands. The spatial distribution of the LBS and VHS states can be observed to be localized only along and across the GB with the increase in $V_{sample}$ from 25 mV → 150 mV → 240 mV → 240 mV in **(d)** → **(e)** → **(f)** → **(g)**.

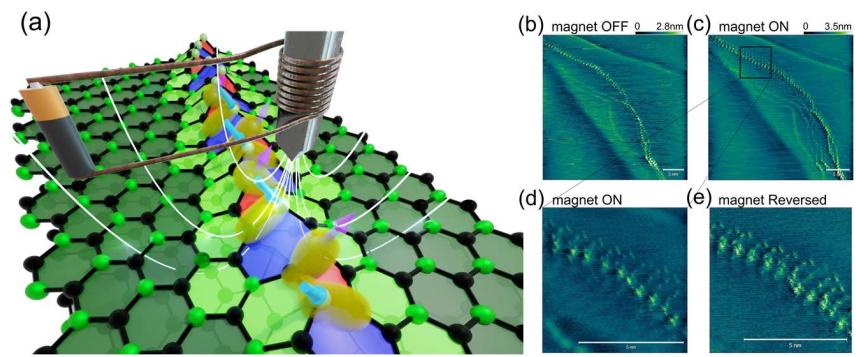

Figure 6: **Spin polarized atomic sensing. (a)** Illustration of atomic spin detection technique on surface features such as GB where external magnetic fields is applied through a current carrying coil wrapped around the tip. The magnetic field can be switched OFF in **(b)** and ON in **(c)**, further magnified in **(d)**, and field-reversed in **(e)**, revealing the spatial distribution of spin-up and spin-down states.

## Supplementary:

### S1: Single GB buckled atoms

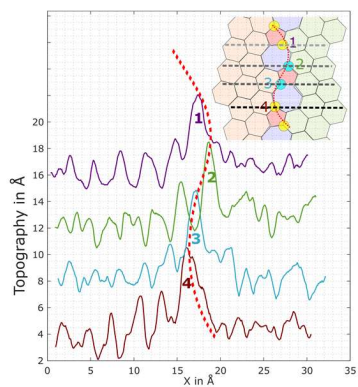

Figure **S1**: The topography across various sites along the buckled GB reveals that only a single GB atom is deformed out-of-plane.

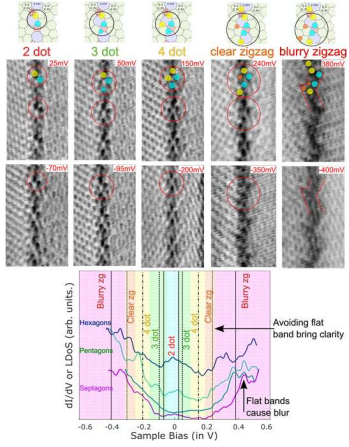

## S2: Positive and negative bias dependent imaging

Figure **S2**: The spatial distribution of the LBS and VHS states can be observed with bias-dependent imaging at positive and negative biases with the increase in $V_{sample}$ from 25 mV → 150 mV → 240 mV. The distribution pattern of states like 2 dot pattern, 3 dot, 4 dot, zigzag and blurry are observed both on positive and negative $V_{sample}$ and are centered around the fermi level.